\documentclass{article}
\usepackage[utf8]{inputenc}
\usepackage{amsmath}
\usepackage{amssymb}
\usepackage{amsfonts}
\usepackage[standard]{ntheorem}
\usepackage{graphicx}

\newcommand{\RR}{\mathbb{R}}

\title{Predicting the Evolution of Gene $ura3$ in the Yeast Saccharomyces Cerevisiae}
\author{Jacques M. Bahi, Christophe Guyeux, and Antoine Perasso}

\begin{document}

\maketitle

\begin{abstract}
Since the late `60s, various genome evolutionary models have been proposed to
predict the evolution of a DNA sequence as the generations pass. 
%Essentially, 
%two main categories of such models can be found in the literature. The first
%one,
Most of these models are based on nucleotides evolution, so they use a mutation matrix of size $4 \times 4$.
They encompass for instance the well-known models of Jukes and Cantor, Kimura, 
and Tamura. 
%In the second category, exclusively studied by Bahi and Michel, 
%the evolution of trinucleotides is studied through a matrix of size 
%$64 \times 64$. 
By essence, all of these models
relate the evolution of DNA sequences to the computation of the successive
powers of a mutation matrix. To make this computation possible, particular
forms for the mutation matrix are assumed, which are not compatible with mutation
rates that have been recently obtained experimentally on gene $ura3$ of the Yeast 
\textit{Saccharomyces cerevisiae}. Using this experimental study, authors of this paper have deduced
a simple mutation matrice, compute the future evolution of the rate 
purine/pyrimidine for $ura3$, investigate the particular 
behavior of cytosines and thymines compared to purines, and simulate the 
evolution of each nucleotide.
\end{abstract}

\section{Introduction}

Codons are not uniformly distributed into the genome.
Over time mutations have introduced some variations in their 
frequency of apparition.
%It can be attractive to study the genetic patterns (blocs of more than one nucleotide: dinucleotides, trinucleotides...) that appear and disappear depending on mutation parameters.
Mathematical models allow the prediction of such an evolution, in such a way 
that statistical values observed into current genomes can be recovered
from hypotheses on past DNA sequences.

A first model for genomes evolution has been proposed in 1969 by Thomas 
Jukes and Charles Cantor \cite{Jukes69}. This first model is very simple,
as it supposes that each nucleotide $A,C,G,T$ has the probability $m$ to
mutate to any other nucleotide, as described in the following mutation
matrix,
$$
\left(\begin{array}{cccc}
* & m & m & m\\
m & * & m & m\\
m & m & * & m\\
m & m & m & *\\
\end{array}\right).
$$
In that matrix, the coefficient in row 3, column 2 represents the 
probability that the nucleotide $G$ mutates in $C$ during the next time
interval, \emph{i.e.}, $P(G \rightarrow C)$. As diagonal elements can be 
deduced by the fact that the sum of each row must be equal to 1, they are
omitted here.

This first attempt has been followed up by Motoo Kimura \cite{Kimura80}, 
who has reasonably considered that transitions ($A \longleftrightarrow G$ and 
$T \longleftrightarrow C$) should not have the same mutation rate than
transversions ($A \longleftrightarrow T$, $A \longleftrightarrow C$, $T 
\longleftrightarrow G$, and $C \longleftrightarrow G$), leading to the
following mutation matrix:
$$
\left(\begin{array}{cccc}
* & b & a & b\\
b & * & b & a\\
a & b & * & b\\
b & a & b & *\\
\end{array}\right).
$$

This model has been refined by Kimura in 1981 (three 
constant parameters, to make a distinction between natural 
$A\longleftrightarrow T$, $C \longleftrightarrow G$ and unnatural
transversions), leading to: 
$$
\left(\begin{array}{cccc}
* & c & a & b\\
c & * & b & a\\
a & b & * & c\\
b & a & c & *\\
\end{array}\right).
$$

Joseph Felsenstein \cite{Felsenstein1980} has then supposed that the nucleotides 
frequency depends on the kind of nucleotide A,C,T,G. Such a supposition leads
to a mutation matrix of the form:
$$
\left(\begin{array}{cccc}
* & \pi_C & \pi_G & \pi_T\\
\pi_A & * & \pi_G & \pi_T\\
\pi_A & \pi_C  & * & \pi_T\\
\pi_A & \pi_C  & \pi_G & *\\
\end{array}\right)
$$
with $3\pi_A, 3\pi_C, 3\pi_G,$ and $3\pi_T$ denoting respectively the frequency of
occurrence of each nucleotide.
Masami Hasegawa, 
Hirohisa Kishino, and Taka-Aki Yano \cite{Hasegawa1985} have generalized the 
models of \cite{Kimura80} and \cite{Felsenstein1980}, introducing in 1985 the
following mutation matrix:
$$
\left(\begin{array}{cccc}
* & \alpha \pi_C & \beta \pi_G & \alpha \pi_T\\
\alpha \pi_A               & * &\alpha \pi_G & \beta\pi_T\\
\beta \pi_A               & \alpha\pi_C  & * & \alpha\pi_T\\
\alpha \pi_A               & \beta\pi_C  & \alpha \pi_G & *\\
\end{array}\right).
$$

These efforts have been continued by Tamura, who proposed in~\cite{Tamura92}
a simple method to estimate the number of nucleotide substitutions per site 
between two DNA sequences, by extending the model of Kimura
(1980). The idea is to consider a two-parameter method, for the case where a 
GC bias exists. Let us denote by $\pi_{GC}$ the frequency of this dinucleotide
motif. Tamura supposes that $\pi_G = \pi_C = \dfrac{\pi_{GC}}{2}$ and
$\pi_A = \pi_T = \dfrac{1-\pi_{GC}}{2}$, which leads to the following rate matrix: 
$$\begin{pmatrix} {*} & {\kappa(1-\pi_{GC})/2} & {(1-\pi_{GC})/2} & {(1-\pi_{GC})/2} \\ {\kappa\pi_{GC}/2} & {*} & {\pi_{GC}/2} & {\pi_{GC}/2} \\ {(1-\pi_{GC})/2} & {(1-\pi_{GC})/2} & {*} & {\kappa(1-\pi_{GC})/2} \\ {\pi_{GC}/2} & {\pi_{GC}/2} & {\kappa\pi_{GC}/2} & {*} \end{pmatrix}.$$

%In the last model of Tamura~\cite{Tamura93}, the two 
%different types of transversions ($A\leftrightarrow T, C\leftrightarrow G$) 
%can have a different rate, whereas transversions 
%are all assumed to occur at the same rate (but that rate is allowed to be 
%different from both of the rates for transitions):
%$$
% \begin{pmatrix} {*} & {\kappa_1\pi_C} & {\pi_A} & {\pi_G} \\ {\kappa_1\pi_T} & {*} & {\pi_A} & {\pi_G} \\ {\pi_T} & {\pi_C} & {*} & {\kappa_2\pi_G} \\ {\pi_T} & {\pi_C} & {\kappa_2\pi_A} & {*} \end{pmatrix}.$$

All these models lead to the so-called GTR model~\cite{yang94}, in which the
mutation matrix has the form (using obvious notations):
$$
\left(\begin{array}{cccc}
* & f_{AC} \pi_C & f_{AG} \pi_G & f_{AT} \pi_T\\
f_{AC} \pi_A               & * & f_{CG} \pi_G & f_{CT} \pi_T\\
f_{AG} \pi_A               & f_{CG} \pi_C  & * & \pi_T\\
f_{AT} \pi_A               & f_{CT}\pi_C  & \pi_G & *\\
\end{array}\right).
$$

Due to mathematical complexity, matrices investigated to model evolution of
DNA sequences are thus limited either by the hypothesis of symmetry or by
the desire to reduce the number of parameters under consideration. These hypotheses
allow their authors to solve theoretically the DNA evolution problem by 
computing directly the successive powers of their mutation matrix. However, one
can wonder whether such restrictions on the mutation rates are realistic. Focusing
on this question, authors of the present paper have used a recent research work 
in which the per-base-pair mutation rates of the Yeast \textit{Saccharomyces cerevisiae}
have been experimentally measured~\cite{Lang08}. Their results are summarized 
in Table~\ref{ura3taux}.

\begin{table}[h]
\centering
 \begin{tabular}{c|c}
Mutation & $ura3$ \\
\hline
$T \rightarrow C$ & 4 \\
$T \rightarrow A$ & 14 \\
$T \rightarrow G$ & 5 \\
$C \rightarrow T$ & 16 \\
$C \rightarrow A$ & 40 \\
$C \rightarrow G$ & 11 \\
$A \rightarrow T$ & 8 \\
$A \rightarrow C$ & 6 \\
$A \rightarrow G$ & 0 \\
$G \rightarrow T$ & 28 \\
$G \rightarrow C$ & 9 \\
$G \rightarrow A$ & 26 \\
Transitions & 46 \\
Transversions & 121 \\
\hline
 \end{tabular}
\caption{Summary of sequenced $ura3$ mutations~\cite{Lang08}}
\label{ura3taux}
\end{table}

%\begin{table}{h}
%\centering
% \begin{tabular}{lcc}
%  \hline
%Mutation & $ura3$ & $CAN1$ \\
%\hline
%$T \rightarrow C$ & 4 & 4\\
%$T \rightarrow A$ & 14 & 9\\
%$T \rightarrow G$ & 5 & 5\\
%$C \rightarrow T$ & 16 & 20\\
%$C \rightarrow A$ & 40 & 21\\
%$C \rightarrow G$ & 11 & 9\\
%$A \rightarrow T$ & 8 & 4\\
%$A \rightarrow C$ & 6 & 5\\
%$A \rightarrow G$ & 0 & 1\\
%$G \rightarrow T$ & 28 & 20\\
%$G \rightarrow C$ & 9 & 12\\
%$G \rightarrow A$ & 26 & 40\\
%Transitions & 46 & 65\\
%Transversions & 121 & 85\\
%\hline
% \end{tabular}
%\caption{Summary of sequenced $ura3$ and $can1$ mutations~\cite{Lang08}}
%\label{ura3taux}
%\end{table}

The mutation matrix of gene $ura3$ can be deduced 
from this table. It is equal to: 

\begin{equation*}
 \left(
\begin{array}{cccc}
 1-m & \dfrac{6m}{14} & 0 & \dfrac{8m}{14}\\
\dfrac{40m}{67} & 1-m & \dfrac{11m}{67} & \dfrac{16m}{67} \\
\dfrac{26m}{63} & \dfrac{9m}{63} & 1-m & \dfrac{28m}{63}\\
\dfrac{14m}{23} & \dfrac{4m}{23} & \dfrac{5m}{23} & 1-m
\end{array}
\right),
\end{equation*}
where $m$ is the mutation rate per generation in $ura3$ gene, which is equal to
$3.80\times 10^{-10}$/bp/generation, or to $3.0552\times 10^{-7}$/generation for
the whole gene~\cite{Lang08}. 
Obviously, none of the existing genomes evolution models can fit such a mutation
matrix, leading to the fact that hypotheses must be relaxed, even if this
relaxation leads to less ambitious models: current models do not
match with what really occurs in concrete genomes, at least in the case of this 
yeast.

Having these considerations in mind, authors of the present article propose to use
the data obtained by Lang and Murray, in order to predict the evolution of 
the rates or purines and pyrimidines in the two genes studied in~\cite{Lang08}.
A mathematical proof giving the intended limit for these rates when the 
generations pass, is reinforced by numerical simulations. The obtained simulations
are thus compared with the historical model of Jukes and Cantor, which is still
used by current prediction software. A model of size $3 \times 3$ with
six independent parameters is then proposed and studied in a case that matches
with data recorded in~\cite{Lang08}. Mathematical investigations and numerical 
simulations  focusing on $ura3$ gene are both given in the case of the yeast 
\textit{Saccharomyces cerevisiae}.

The remainder of this research work is organized as follows.
In Sections \ref{Model22} and \ref{Model33}, we focus on the evolution of the gene $ura3$. Section \ref{Model22} is dedicated to the formulation of a non symmetric discrete model of size $2 \times 2$. This model translates a genome evolution taking into account purines and pyrimidines mutations. A simulation is then performed to compare this non symmetric model to the classical symmetric Cantor model. Section \ref{Model33} deals with a 6-parameters non symmetric model of size $3 \times 3$, focusing on the one hand on the evolution of purines and on the other hand of cytosines and thymines. This mathematical model is illustrated throughout simulations of the evolution of the purines, cytosines and thymines of gene $ura3$. We finally conclude this work in Section \ref{Conclusion}.



%% References


\section{Non-symmetric Model of size $2\times 2$}
\label{Model22}

In this section, a first general genome evolution model focusing on purines
versus pyrimidines is proposed, to illustrate the method and as a pattern
for further investigations. This model is applied to the case of the yeast
\textit{Saccharomyces cerevisiae}.

\subsection{Theoretical Study}

Let $R$ and $Y$ denote respectively the occurrence frequency of purines and pyrimidines in a sequence
of nucleotides, and $M=\left(
\begin{array}{cc}
	a & b \\
	c & d
\end{array}\right)
$
the associated mutation matrix, with
$a = P(R\to R)$, $b=P(R\to Y)$, $c=P(Y\to R)$, and $d=P(Y\to Y)$
satisfying
\begin{equation}\label{stochastic}
\begin{cases}
a+b = 1, \\
c+d = 1, \\
%1\geq a>b \geq 0,\\
%1\geq c>d\geq 0,  % C'est faux
\end{cases}
\end{equation}
and thus $M=\left(
\begin{array}{cc}
	a & 1-a \\
	c & 1-c
\end{array}\right)
$.

The initial probability is denoted by $P_0 = (R_0 ~~ Y_0)$, where $R_0$ and $Y_0$ denote
respectively the initial frequency of purines and pyrimidines. So the occurrence 
probability at generation $n$ is $P_n =  P_0 M^n$, where $P_n=(R(n) ~~ Y(n))$ 
is a probability vector such that $R(n)$ (resp. $Y(n)$) is the rate of purines
(resp. pyrimidines) after $n$ generations.
\vspace{0.5cm}

\underline{Determination of $M^n$}

A division algorithm leads to the existence of a polynomial of degree $n-2$,
denoted by $Q_M \in \RR_{n-2}[X]$, and to $a_n,b_n\in\RR$ such that
\begin{equation}\label{div}
X^n = Q_M(X) \chi_M(X) + a_n X + b_n,
\end{equation}
when $\chi_M$ is the characteristic polynomial of $M$. 
Using both the Cayley-Hamilton theorem and the 
equality given above, we thus have
\begin{equation*}
M^n = a_n M + b_n I_2.
\end{equation*}

In order to determine $a_n$ and $b_n$, we must find the roots of $\chi_M$. As $\chi_M(X) = X^2- Tr(M) X + \det(M)$ and due to \eqref{stochastic},  we can conclude that $1$ is a root of $\chi_M$, which thus has two real roots: $1$ and $x_2$. 
As the roots sum is equal to -tr(A), we conclude that $x_2 = a-c$.

If $x_2 = a-c = 1$, then $a=1$ and $c=0$ (as these parameters are in $[0,1]$),
 so the mutation matrix is the identity and the frequencies of purines and pyrimidines
into the DNA sequence does not evolve. If not, evaluating  \eqref{div} in both $X=1$ and $X=x_2$, we thus obtain
\begin{equation*}
\begin{cases}
1 = a_n + b_n, \\
(a-c)^n = a_n (a-c) + b_n.
\end{cases}
\end{equation*}
Considering that $a-c \neq 1$, we obtain
\begin{equation*}
a_n  = \frac{(a-c)^n - 1}{a-c-1}, \qquad b_n = \frac{a-c - (a-c)^n}{a-c-1}.
\end{equation*}
Using these last expressions into the equality 
linking $M$, $a_n$, and $b_n$, we thus deduce the value of $P_n = P_0 M^n$, where
\begin{equation}
\label{mn2d}
M^n = \frac{1}{a-c-1} \left(
\begin{array}{cc}
	(a-1) (a-c)^n - c & (1-a) ((a-c)^n -1)) \\
	c ((a-c)^n-1) & -c (a-c)^n + a -1
\end{array}\right).
\end{equation}

\begin{figure}
\centering
\includegraphics[scale=0.5]{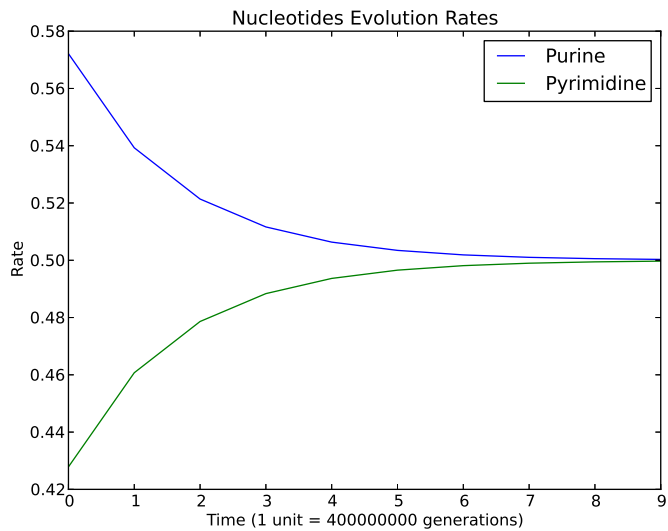}
\caption{Prediction of purine/pyrimidine evolution of \emph{ura3} gene in symmetric Cantor model.}
\label{Cantor2d}
\end{figure}

If $a=0$ and $c=1$, then 
$M = \left(\begin{array}{cc}0 & 1 \\ 1 & 0\end{array}\right)$, so $M^{2n}$ is
the identity $I_2$ whereas $M^{2n+1}$ is $M$. Contrarily, if 
$(a,c) \notin \{(0,1); (1,0)\}$, then the limit of $M^n$ can be easily 
found using \eqref{mn2d}, leading to the following result.

\begin{theorem}
\label{th2d}
Consider a DNA sequence under evolution, whose mutation matrix is
$M=\left(
\begin{array}{cc}
	a & 1-a \\
	c & 1-c
\end{array}\right)
$
with $a=P(R\to R)$ and $c = P(Y\to R)$. 
\begin{itemize}
 \item If $a=1, c=0$, then the frequencies of purines and pyrimidines do not 
change as the generation pass.
 \item If $a=0, c=1$, then these frequencies oscillate at each generation between 
$(R_0 ~~ Y_0)$ (even generations) and $(Y_0 ~~ R_0)$ (odd generations).
 \item Else the value $P_n = (R(n) ~~ Y(n))$ of purines and pyrimidines 
frequencies at generation $n$ is convergent to the following limit:
\begin{equation*}
\lim_{n\to\infty} P_n  = \frac{1}{c+1-a}  \left(
\begin{array}{cc}
	c &	1-a
\end{array}\right).
\end{equation*}
\end{itemize}
\end{theorem}

\subsection{Numerical Application}

For numerical application, we will consider mutations rates
in the \emph{ura3} gene of the Yeast \emph{Saccharomyces cerevisiae}, as obtained by Gregory I. Lang and Andrew W.
Murray~\cite{Lang08}. As stated before, they have measured phenotypic 
mutation rates, indicating that the per-base pair mutation
rate at \emph{ura3} is equal to $m=3.0552\times 10^{-7}$/generation.
For the majority of Yeasts they studied, \emph{ura3} is constituted by 804 bp: 133 cytosines,
211 thymines, 246 adenines, and 214 guanines. So $R_0 = \dfrac{246+214}{804} \approx 0.572$,
and $Y_0 = \dfrac{133+211}{804} \approx 0.428$. 
Using these values in the historical model of Jukes and Cantor~\cite{Jukes69}, we obtain the evolution depicted in
Figure~\ref{Cantor2d}. %, whereas the first Kimura leads to simulations depicted
%in Fig.~\ref{Kimura2d}. 

% 
% \begin{color}{red}
% Si on a le temps pour une g\'en\'eration, on peut pr\'evoir quand on 
% aura exactement autant de purines que de pyrimidines.
% \end{color}
% 

% 
% \begin{figure}
% \centering
% \includegraphics[scale=0.5]{Kimura1_2D.eps}
% \caption{Prediction of purine/pyrimidine evolution of \emph{ura3} gene in symmetric Cantor model.}
% \label{Kimura2d}
% \end{figure}
% 

\begin{figure}
\centering
\includegraphics[scale=0.5]{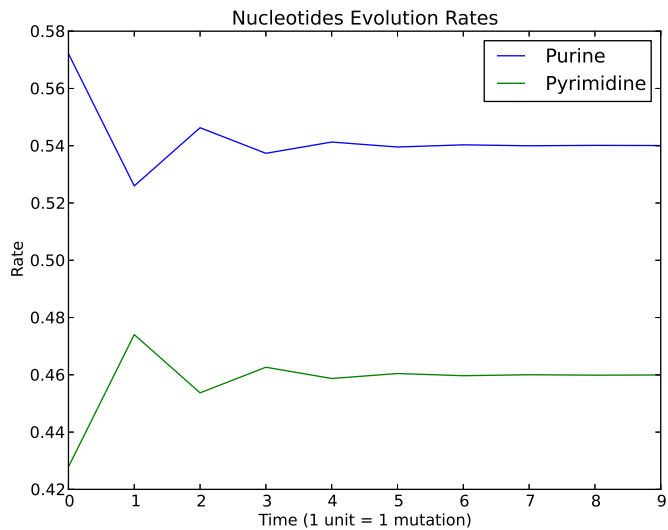}
\caption{Prediction of purine/pyrimidine evolution of \emph{ura3} gene in non-symmetric Model of size $2\times 2$.}
\label{evolution par notre methode}
\end{figure}

Theorem \ref{th2d} allow us to compute the limit of the rates of purines and 
pyrimidines:
\begin{description}
 \item[Computation of probability $a$.]
$a = P(R \rightarrow R) = (1-m)+ P(A \rightarrow G) + P(G \rightarrow A)$. The use of Table~\ref{ura3taux} implies
that $a = (1-m) + m \left(\dfrac{0+26}{x}\right)$,
where $x$ is such that $1-a = P(R \rightarrow Y)
= m\left(\dfrac{6+8+28+9}{x}\right)$, \emph{i.e.},
$x=77$, and so $a=1-\dfrac{51m}{77}$.
 \item[Computation of probability $c$.]
Similarly, $c = P(Y \rightarrow R) 
= P(C \rightarrow A) + P(C\rightarrow G) + P(T \rightarrow A) 
+ P(T\rightarrow G) = \dfrac{70m}{y}$, whereas
$1-c = 1-m+\dfrac{20m}{y}$.
So $c = \dfrac{7m}{9}$.
\end{description}
 
The purine/pyrimidine mutation matrix that 
corresponds to the data of~\cite{Lang08} is thus
equal to:
$$M=\left(
\begin{array}{cc}
	1-\dfrac{51m}{77} & \dfrac{51m}{77} \\
	\dfrac{7m}{9} & 1-\dfrac{7m}{9}
\end{array}\right).
$$

Using the value of $m$ for the $ura3$ gene leads
to $1-a=2.02357\times 10^{-7}$ and $c=2.37627\times 10^{-7}$,
which can be used in Theorem~\ref{th2d} to 
conclude that the rate of pyrimidines is convergent
to $45.992\%$ whereas the rate of purines 
converge to $54.008\%$.
Numerical simulations using data published 
in~\cite{Lang08} are given in
Figure~\ref{evolution par notre methode}, leading
to a similar conclusion.

\section{A First Non-Symmetric Genomes Evolution Model of size $3 \times 3$ having 6 Parameters}
\label{Model33}

In order to investigate the evolution of the frequencies of cytosines and thymines
in the gene $ura3$, a model of size $3 \times 3$ compatible with real
mutation rates of the yeast \textit{Saccharomyces cerevisiae} is now presented.

\subsection{Formalization}

Let us consider a line of yeasts where a given gene is sequenced at
each generation, in order to clarify explanations. The $n-$th generation
is obtained at time $n$, and the rates of purines, cytosines, and tymines
at time $n$ are respectively denoted by $P_R(n), P_C(n)$, and $P_T(n)$.

Let $a$ be the probability that a purine is changed into a cytosine between
two generations, that is:  $a = P(R \rightarrow C)$. 
Similarly, denote by $b,c,d,e,f$ the 
respective probabilities: $P(R \rightarrow T)$, $P(C \rightarrow R)$, 
$P(C \rightarrow T)$, 
$P(T \rightarrow R)$, and $P(T \rightarrow C)$. Contrary to existing 
approaches, $P(R \rightarrow C)$ is not supposed to be equal to 
$P(C \rightarrow R)$, and the
same statement holds for the other probabilities. For the sake of simplicity,
we will consider in this first research work that $a,b,c,d,e,f$ are not 
time dependent.

Let 
\begin{equation*}
 M = \left( 
\begin{array}{ccc}
1-a-b & a & b \\
c & 1-c-d & d \\
e & f & 1-e-f
\end{array}
\right)
\end{equation*}
be the mutation matrix associated to the probabilities mentioned above, and $P_n$ 
the vector of occurrence, at time $n$, of each of the three kind of
nucleotides. In other words, 
$P_n = (P_R(n) ~~ P_C(n) ~~ P_T(n))$. 
Under that hypothesis, $P_n$ is a 
probability vector: $\forall n \in \mathbb{N},$
\begin{itemize}
 \item $P_R(n), P_C(n), P_T(n) \in [0,1]$,
 \item $P_R(n) + P_C(n) + P_T(n) = 1$,
\end{itemize}
Let $P_0 = (P_R(0) ~~ P_C(0) ~~ P_T(0)) \in [0,1]^3$ be the initial probability vector. We have 
obviously: 
\begin{equation*}
P_R(n+1) = P_R(n) P(R \rightarrow R)+
P_C(n) P(C \rightarrow R)+P_T(n) P(T \rightarrow R). 
\end{equation*}
Similarly,
$P_C(n+1) = P_R(n) P(R \rightarrow C)+
P_C(n) P(C \rightarrow C)+P_T(n) P(T \rightarrow C)$ and
$P_T(n+1) = P_R(n) P(R \rightarrow T)+
P_C(n) P(C \rightarrow T)+P_T(n) P(T \rightarrow T)$. 
This equality yields the following one, 
\begin{equation}
\label{Pn}
 P_n = P_{n-1} M = P_0 M^n.
\end{equation}
In all that follows we wonder if, given the parameters $a,b,c,d,e,f$ as in~\cite{Lang08},
one can determine the frequency of occurrence of any of the three kind
of nucleotides when $n$ is sufficiently large, in other words if the limit of
$P_n$ is accessible by computations.

\subsection{Resolution}

%\subsubsection{Determination of $M^n$ in the general case}

The characteristic polynomial of $M$ is equal to
\begin{equation*}
\begin{array}{cl} 
\chi_M(x) &= x^3 +(s-3) x^2+(p-2s+3)x-1+s-p \\
 & = (x-1)\left(x^2+(s-2)x+(1-s+p)\right),
\end{array}
\end{equation*}
where
\begin{gather*}
s = a+b+c+d+e+f, \\
p=ad+ae+af+bc+bd+bf+ce+cf+de, \\
det(M)  =1-s+p.
\end{gather*}

The discriminant of the polynomial of degree 2 in the factorization of $\chi_M$ is equal to 
$\Delta = (s-2)^2-4(1-s-p) = s^2-4p$. Let $x_1$ and
$x_2$ the two roots (potentially complex or 
equal) of $\chi_M$, given by
\begin{equation}
\label{x1x2}
x_1 = \dfrac{-s+2-\sqrt{s^2-4p}}{2} \textrm{ and }  x_2 = \dfrac{-s+2+\sqrt{s^2-4p}}{2}. 
\end{equation}

Let $n \in \mathbb{N}, n \geqslant 2$. As $\chi_M$ is a polynomial of degree 3, a division algorithm of $X^n$ by $\chi_M(X)$ leads to the
existence and uniqueness of two polynomials $Q_n$ and $R_n$, such that
\begin{equation}
\label{resol}
X^n = Q_n(X)\chi_2(X)+R_n(X),
\end{equation}
where the degree of $R_n$ is lower than or equal to
the degree of $\chi_M$, \emph{i.e.}, 
$R_n(X) = a_n X^2 + b_n X + c_n$ with $a_n, b_n, c_n \in \mathbb{R}$ for every $n\in\mathbb{N}$.
By evaluating \eqref{resol} in the three
roots of $\chi_M$, we find the system
$$\left\{ 
\begin{array}{cl}
1 & = a_n + b_n + c_n\\
x_1^n & = a_n x_1^2 + b_n x_1 + c_n\\
x_2^n & = a_n x_2^2 + b_n x_2 + c_n\\
\end{array}
\right.$$
This system is equivalent to
$$\left\{ 
\begin{array}{cccccl}
c_n & + &  b_n  &+ & a_n & = 1\\
    && b_n(x_1-1)& + & a_n(x_1^2-1) & = x_1^n-1\\
    && b_n(x_2-1) &+ & a_n(x_2^2-1) &  = x_2^n-1\\ 
\end{array}
\right.$$
For the $ura3$ gene, it is easy to check that $x_1 \neq 1$, $x_2 \neq 1$, 
and $x_1 \neq x_2$ (see numerical applications of Section~\ref{appnum}).
Then standard algebraic computations give
\begin{equation*}
\left\{
\begin{array}{l}
a_n = \dfrac{1}{x_2-x_1}\left[\dfrac{x_2^n-1}{x_2-1}-\dfrac{x_1^n-1}{x_1-1}\right],\\
\\
b_n = \dfrac{x_1+1}{x_1-x_2}\dfrac{x_2^n-1}{x_2-1} + \dfrac{x_2+1}{x_2-x_1}\dfrac{x_1^n-1}{x_1-1},\\
\\
c_n=1-a_n-b_n.
\end{array}
\right.
\end{equation*}
Using for $i=1,2$ and $n\in\mathbb{N}$ the following notation,
\begin{equation}\label{Xi}
X_i(n) = \dfrac{x_i^n-1}{x_i-1}, 
\end{equation}
and since $x_2-x_1=\sqrt{\Delta}$, the system above
can be rewritten as
\begin{equation}
\label{soluce}
\left\{
\begin{array}{l}
a_n = \dfrac{X_2(n)-X_1(n)}{\sqrt{\Delta}},\\
\\
b_n = \dfrac{(x_2+1)X_1(n)-(x_1+1)X_2(n)}{\sqrt{\Delta}},\\
\\
c_n=1+\dfrac{x_1 X_2(n)-x_2 X_1(n)}{\sqrt{\Delta}}.
\end{array}
\right.
\end{equation}
By evaluating \eqref{resol} in $M$ and due to the
theorem of Cayley-Hamilton, we finally have for every integer $n\geqslant 1$,
\begin{equation}
\label{Mn}
M^n = a_n M^2 + b_n M + c_n I_3,
\end{equation}
where $I_3$ is the identity matrix of size 3, 
$a_n, b_n,$ and $c_n$ are given by \eqref{soluce}, and $M^2$ is given by
$$
M^2=\left(
\begin{array}{c|c|c}
a^2 + 2ab + ac - 2a & -a^2 - ab - ac  & -ab + ad - b^2 \\
+ b^2 + be - 2b + 1  & - ad + 2a + bf & - be - bf + 2b\\
\hline
-ac - bc - c^2  & ac + c^2 + 2cd - 2c  & bc - cd - d^2 \\
- cd + 2c + de & + d^2 +df - 2d + 1 & - de - df + 2d\\
\hline
-ae - be + cf  & ae - cf - df  & be +df + e^2 + 2ef \\
- e^2 - ef + 2e & - ef - f^2 + 2f & - 2e + f^2 - 2f + 1
\end{array}
\right)
.$$

\subsection{Convergence study}

In the case of $ura3$, $|x_1|<1$ and $|x_2|<1$ (see the next section). Then $X_i(n) \longrightarrow \dfrac{1}{1-x_i}$ for
$i=1,2$ and so
$$a_n \longrightarrow \dfrac{1}{\sqrt{\Delta}}\left(\dfrac{1}{1-x_2}-\dfrac{1}{1-x_1}\right).$$
Denote by $a_\infty$ this limit. We have 
$$a_\infty = \dfrac{x_2-x_1}{\sqrt{\Delta}(1-x_2)(1-x_1)} = \dfrac{1}{(1-x_2)(1-x_1)} = \dfrac{1}{\dfrac{s+\sqrt{\Delta}}{2}\dfrac{s-\sqrt{\Delta}}{2}},$$
and finally
\begin{equation*}
a_\infty = \dfrac{4}{s^2-\Delta}=\dfrac{1}{p}. 
\end{equation*}
Similarly, $b_n = X_1(n)-a_n (x_1+1)$ satisfies 
\begin{equation*}
b_n\longrightarrow \dfrac{1}{1-x_1} - \dfrac{x_1+1}{p}.
\end{equation*}
The following computations
\begin{gather*}
\dfrac{1}{1-x_1} = \dfrac{2}{s+\sqrt{\Delta}} = \dfrac{2(s-\sqrt{\Delta})}{s^2-\Delta} = \dfrac{s-\sqrt{\Delta}}{2p}, \\
\dfrac{x_1+1}{p} = \dfrac{-s+4-\sqrt{\Delta}}{2p},
\end{gather*}
finally yield
\begin{equation*}
b_\infty = \dfrac{s-2}{p}.
\end{equation*}
So  
\begin{equation*}
c_n \longrightarrow 1-a_\infty-b_\infty = \dfrac{p-s+1}{p},
\end{equation*}
and to sum up, the distribution limit is given by
\begin{equation}
%\label{soluce}
\left\{
\begin{array}{l}
a_\infty = \dfrac{1}{p}\\
\\
b_\infty = \dfrac{s-2}{p}\\
\\
c_\infty=\dfrac{p-s+1}{p}
\end{array}
\right.
\end{equation}
Using the latter values in \eqref{Mn}, we can 
determine the limit of $M^n$, which is
$a_\infty M^2+b_\infty M + c_\infty I_3$.
All computations done, we find the following limit for $M^n$,
\begin{equation*}
\dfrac{1}{p-bf+df}\left(\begin{array}{ccc}
ce+cf+de-bf+df & ae+af+bf & ad+bc+bd \\            
ce+cf+de & ae+af+df & ad+bc+bd \\                   
ce+cf+de & ae+af+bf & ad+bc+bd-bf+df\\                   
                  \end{array}
\right).
\end{equation*}
Using \eqref{Pn}, we can thus finally determine
the limit of $P_n = P_0 M^n$ \linebreak $= 
(P_R(0) ~~ P_C(0) ~~ P_T(0)) M^n $, which leads to
the following result.
\begin{theorem}
\label{th1x}
The frequencies $P_R(n), P_C(n)$, and
$P_T(n)$ of occurrence at
time $n$ of purines, cytosines, and thymines in the considered gene $ura3$ of
the yeast \textit{Saccharomyces cerevisiae}
 converge to the following values:
\begin{itemize}
 \item $P_R(n)\longrightarrow \dfrac{ce+cf+de + (df-bf) P_R(0)}{p-bf+df}$
 \item $P_C(n)\longrightarrow \dfrac{ae+af+df + (df-bf) P_C(0)}{p-bf+df}$
 \item $P_T(n)\longrightarrow \dfrac{ad+bc+bd + (df-bf) P_T(0)}{p-bf+df}$
\end{itemize}
\end{theorem}

\subsection{Numerical Application and Simulations}
\label{appnum}

We consider another time the numerical values for mutations published 
in~\cite{Lang08}. Gene \emph{ura3} of the Yeast \textit{Saccharomyces cerevisiae} has a mutation rate of
$3.80 \times 10^{-10}$/bp/generation~\cite{Lang08}. As this gene is constituted by 804
nucleotides, we can deduce that its global mutation rate per generation is equal to
$m = 3.80 \times 10^{-10} \times 804 = 3.0552\times 10^{-7}$.
Let us compute the values of $a,b,c,d,e,$ and $f$.
The first line of the mutation matrix is constituted by 
$1-a-b = P(R \rightarrow R)$, $a=P(R \rightarrow T)$,
and $b=P(R \rightarrow C)$. $P(R \rightarrow R)$ takes into account
the fact that a purine can either be preserved (no mutation, probability $1-m$), 
or mutate into another purine ($A \rightarrow G$, $G \rightarrow A$).
As the generations pass, authors of~\cite{Lang08} have counted 0 mutations of
kind $A \rightarrow G$, and 26 mutations of kind $G \rightarrow A$.
Similarly, there were 28  mutations $G \rightarrow T$ and 8: $A \rightarrow T$,
so 36: $R \rightarrow T$. Finally, 6: $A \rightarrow C$ and 9: $G \rightarrow C$ 
lead
to 15: $R \rightarrow C$ mutations. The total of mutations to consider when 
evaluating the first line is so equal to 77. All these considerations lead
to the fact that $1-a-b=(1-m)+m\dfrac{26}{77}$, $a=\dfrac{36m}{77}$, and $b=\dfrac{15m}{77}$. A similar reasoning leads to $c=\dfrac{19m}{23}$, $d=\dfrac{4m}{23}$, 
$e=\dfrac{51m}{67}$, and $f=\dfrac{16m}{67}$. 

In that situation, $s=a+b+c+d+e+f=\dfrac{205m}{77} \approx 8.134\times 10^{-7}$, and $p= \dfrac{207488 m^2}{118657}
\approx 1.632\times 10^{-13}.$
So $\Delta = s^2-4p = \dfrac{854221 m^2}{9136589} >0$, $x_1 = 1-\dfrac{m}{2}\left(\dfrac{205}{77}+\sqrt{\dfrac{854221}{9136589}}\right)$, and $x_2 = 1-\dfrac{m}{2}\left(\dfrac{205}{77}-\sqrt{\dfrac{854221}{9136589}}\right)$.
As $x_1\approx 0.9999685 \in [0,1]$ and $x_2 \approx 0.9999686 \in [0,1]$, we have, due to Theorem~\ref{th1x}:
\begin{itemize}
 \item $P_R(n)\longrightarrow \dfrac{ce+cf+de + (df-bf) P_R(0)}{p-bf+df}$
 \item $P_C(n)\longrightarrow \dfrac{ae+af+df + (df-bf) P_C(0)}{p-bf+df}$
 \item $P_T(n)\longrightarrow \dfrac{ad+bc+bd + (df-bf) P_T(0)}{p-bf+df}$
\end{itemize}

\begin{figure}
\centering
\includegraphics[scale=0.5]{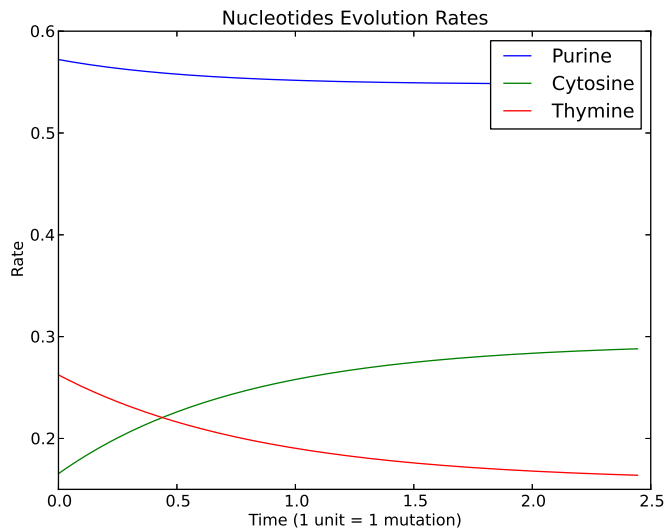}
\caption{Prediction of evolution concerning the purine, thymine, and cytosine rates in \emph{ura3}. Non-symmetric Model of size $3\times 3$.}
\label{3param}
\end{figure}

Using the data of~\cite{Lang08}, we find that $P_R(0)=\dfrac{460}{804}\approx 0.572$,
$P_C(0)=\dfrac{133}{804}\approx 0.165$, and
$P_T(0)=\dfrac{211}{804}\approx 0.263$. So $P_R(n) \longrightarrow 0.549$, $P_C(n) \longrightarrow 0.292$, and 
$P_T(n) \longrightarrow 0.159$.
Simulations corresponding to this example are given in Fig.~\ref{3param}.

\section{Conclusion}
\label{Conclusion}

In this document, the possible evolution of gene $ura3$ of the yeast 
\textit{Saccharomyces cerevisiae} has been studied. As current models of 
nucleotides cannot fit the mutations obtained experimentally by Lang and 
Murray~\cite{Lang08}, authors of this paper have introduced two new simple
models to predict the evolution of this gene. 
On the one hand, a formulation of a non symmetric discrete model of size 
$2 \times 2$ has been proposed, which studies a DNA evolution taking into 
account purines and pyrimidines mutation rates. A simulation has been performed, 
to compare the proposal to the well known Jukes and
Cantor model. On the other hand, a 6-parameters non symmetric 
model of size $3 \times 3$ has been introduced and tested with numerical 
simulations, to make a distinction 
between cytosines and thymines in the former proposal.
These two models still remain generic, and can be adapted to a large panel of
applications, replacing either the couple (purines, pyrimidines) or the tuple
(purines, cytosines, thymines) by any categories of interest.

The $ura3$ gene is not the unique example of a DNA sequence of interest such that none of
the existing nucleotides evolution models cannot be applied due to a complex
mutation matrix. For instance, a second gene called $can1$ has been studied too
by the authors of~\cite{Lang08}. Similarly to gene $ura3$, usual models cannot be used to predict the
evolution of $can1$, whereas a study following a same canvas than what has 
been proposed in this research work can be realized.
In future work, the authors' intention is to make a complete mathematical
study of the 6-parameters non symmetric  model of size $3 \times 3$ proposed
in this document, and to apply it to various case studies. Biological consequences
of the results produces by this model will be systematically investigated. Then,
the most general non symmetric model of size $4 \time 4$ will be regarded in some
particular cases taken from biological case studies, and the possibility of 
mutations non uniformly distributed will then be investigated.

%\begin{table}[h]
%\centering
% \begin{tabular}{lc}
%  \hline
%Mutation & $can1$  \\
%\hline
%$T \rightarrow C$ & 4 \\
%$T \rightarrow A$ & 9 \\
%$T \rightarrow G$ & 5 \\
%$C \rightarrow T$ & 20 \\
%$C \rightarrow A$ & 21 \\
%$C \rightarrow G$ & 9 \\
%$A \rightarrow T$ & 4 \\
%$A \rightarrow C$ & 5 \\
%$A \rightarrow G$ & 1 \\
%$G \rightarrow T$ & 20 \\
%$G \rightarrow C$ & 12 \\
%$G \rightarrow A$ & 40 \\
%Transitions & 65 \\
%Transversions & 85 \\
%\hline
% \end{tabular}
%\caption{Summary of sequenced $can1$ mutations~\cite{Lang08}}
%\label{can1table}
%\end{table}

\bibliographystyle{plain}
\bibliography{mabase}

\end{document}